\documentclass[onecolumn,fleqn,usenatbib]{mnras}
\usepackage[T1]{fontenc}
\usepackage{ae,aecompl}
\usepackage{amsfonts}
\usepackage{amssymb}
\usepackage{amsmath}
\usepackage{graphicx}
\usepackage{epstopdf}
\usepackage[toc,page]{appendix}
\usepackage{float}
\usepackage{soul}

\newcommand{\vel}{\boldsymbol{{u}}}

\newcommand{\B}{\boldsymbol{{B}}}
\newcommand{\A}{\boldsymbol{{A}}}

\title{On the dynamical interaction between  overshooting convection and an underlying dipole
magnetic field - I. The non-dynamo  regime}

\author[L. Korre et al.]{L. Korre$^{1}$\thanks{E-mail:lydia.korre@lasp.colorado.edu},
N.H. Brummell$^{2}$,
P. Garaud$^{2}$, and
C. Guervilly$^{3}$
\\
$^{1}$
Laboratory for Atmospheric and Space Physics,  Boulder, Colorado 80303, USA \\
$^{2}$
Department of Applied Mathematics, Jack Baskin School of Engineering, 
University of California Santa Cruz, \\ Santa Cruz, California 95064, USA \\
$^{3}$
School of Mathematics, Statistics and Physics, Newcastle University,
Newcastle upon Tyne, NE17RU, UK}
\date{Accepted XXX. Received YYY; in original form ZZZ}

\pubyear{2021}

\begin{document}
\label{firstpage}
\pagerange{\pageref{firstpage}--\pageref{lastpage}}
\maketitle
\begin{abstract}
Motivated by the dynamics in the deep interiors of many stars, we study the interaction between overshooting convection and the large-scale poloidal fields residing in radiative zones.  We have run a suite of 3D Boussinesq numerical calculations in a spherical shell that consists of a convection zone with an underlying stable region  that  initially compactly contains a dipole  field. By varying the strength of the convective driving, we find that, in the less turbulent regime,  convection acts as turbulent diffusion that removes the field faster than solely molecular  diffusion would do. However, in the more turbulent regime, turbulent pumping becomes more efficient and partially counteracts turbulent diffusion,  leading to a local accumulation of the field below the overshoot region. These simulations suggest that dipole fields might be  confined in underlying stable regions by highly turbulent convective motions at  stellar parameters.  The confinement is of large-scale field in an average sense and we show that it is reasonably modeled by mean-field ideas. Our findings are  particularly interesting for certain models of the Sun, which  require a large-scale, poloidal magnetic field to be confined in the solar radiative zone in order to explain simultaneously  the uniform rotation of the latter and  the thinness of the solar tachocline. 
\end{abstract} 

\section{Introduction}
\label{sec:intro}
The interface between the outer convection zone and inner radiative zone of solar-type stars is a region of crucial importance for stellar evolution in terms of its dynamical role in the transport of chemical species and angular momentum, and for the generation of magnetic fields. In this series of papers, we investigate this region in detail using numerical simulations. Our previous paper \citep*{Korre19} concentrated on purely hydrodynamic processes connecting these two zones, and here we turn to the study of magnetohydrodynamic processes. \par Magnetism is arguably the most readily dismissed aspect of stellar astrophysics, despite the fact that magnetic fields are undeniably ubiquitous at all stellar masses and all stages of stellar evolution \citep[see][]{Mestel99}.  Magnetic fields are expected to be found everywhere in a given star, from the core to the surface, and in both radiative and convective regions. In the highly turbulent convection zones of solar-type stars, even the weakest seed field can be amplified through dynamo action, up to amplitudes that are dynamically significant and often observable using various techniques including starspot tracking and Zeeman  imaging, for instance \citep[see the review by][]{DL2009}. In the far more quiescent underlying radiation zones, by contrast, any resident magnetic field is usually thought to be of primordial origin  or to originate from a nearby convective region through advection or diffusion (\citealp[see][]{Garaud99,Tobias2001}; \citealp[although  for alternative theories see also][]{Spruit99}). Regardless of their origin, radiation zone fields are usually thought to be  large-scale, because small-scale fields would decay on a time-scale that is short compared with the stellar evolution time-scale. \par The main impact of large-scale magnetic fields in stellar radiative zones comes from their ability to transport angular momentum very efficiently through magnetic stresses. In ideal MHD (i.e. in a plasma where the magnetic diffusivity $\eta$ is zero), Ferraro's law of isorotation \citep{Ferraro} states that the rotation rate of a fluid is forced to be constant along magnetic field lines. Although stellar plasmas have finite diffusivity, the latter is very small in stellar radiation zones, and Ferraro's isorotation law applies even if the field amplitude is very small.  As demonstrated by \citet{MW1987}, a field as low as a few mG is in principle capable of imposing a roughly uniform rotation in the star's radiative zone. 
Ferraro's isorotation law has important observable consequences for the evolution of the angular velocity profile within solar-type stars undergoing magnetic spin-down. \citet{CMG1993} studied this problem and demonstrated that the overall response of the star's rotation profile to the spin-down depends crucially on whether an assumed large-scale poloidal field $\B_p$ is entirely confined to the radiative interior (i.e. such that $\B_p= 0$ in the convective zone), or whether these field lines extend into the overlying convection zone (i.e. such that $\B_p \ne 0$ in the convective zone). In the former case, each zone ultimately ends up rotating almost uniformly, but with different rotation rates. In the latter case, the radiative and convective zones are magnetically connected and the entire star rotates almost uniformly. The rotation profile within the star thus strongly depends on the field's geometry (confined vs. unconfined). 
\par There is a wealth of observational evidence for both types of configurations in stars of all ages. Observations of the rotation rates of solar-type stars in a$~100$Myr-old stellar cluster \citep{Irwin2007} show that a two-zone model is required for slowly-rotating stars in the mass range $0.7M_\odot - 1.1 M_\odot$, suggesting they might have a confined field structure. Rapid rotators in the same mass range are consistent with uniform rotation, by contrast, suggesting an open field structure.
The Sun has been observed (through helioseismology) to possess an almost uniformly rotating inner radiative zone, whereas the outer convection zone rotates differentially with a faster equator and slower poles \citep{seismic1996}. The two zones are separated by a thin shear layer called the solar tachocline. \citet{GM98} argued that this could only be explained by the presence of an embedded primordial magnetic field, strictly confined below the base of the convection zone.  In red giant branch (RGB) stars, finally, the ratio of the core to envelope rotation rates observed using asteroseismology is far lower than what one would expect from angular momentum conservation only (\citealp{RGB, Mosser2012, Marques13}; \citealp*[for a review, see also][]{Aerts19}). Angular momentum transport by a combination of large-scale magnetic fields and large-scale flows could provide an explanation for the increased dynamical coupling between the core and the envelope, but would require a confined field structure, as in the \citet{GM98}  model \citep[see, e.g.][]{OG13}.  
\par In all examples described above, however, the question of {\it how} and {\it when} a poloidal magnetic field might be confined is still essentially unanswered.   For example, \citet{CMG1993}, \citet{RK97} and \citet{MGC1999} assumed  a given confined poloidal field shape, and only investigated the dynamical interaction between this field and the angular velocity profile of the star.  \citet{GM98} were the first to model the confinement of the magnetic field self-consistently, but their model was two-dimensional, and took the form of a steady-state boundary layer analysis. Magnetic confinement in their paper was defined as the ability of the meridional flows to advect the poloidal field downward in a way that compensates the diffusion of the field outward. As a result, the amplitude of the poloidal field was found to decrease exponentially away from this ``advection-diffusion" layer. This idea was essentially confirmed by \citet*{AG13}, who found steady-state nonlinear numerical solutions of the \citet{GM98} model exhibiting confinement \citep*[see also][]{WMc2011,WMG2011}. 
\par Until the early 2000s, theoretical ideas associated with magnetic confinement primarily revolved around slow, laminar flows, and the question of the role of fast turbulent processes naturally arose (especially as the ability to model 3D time-dependent MHD through super-computing became more prevalent). To understand why, it is useful to split the field and the flow into large scales ($\langle\vel\rangle$ and $\langle\B\rangle$) and fluctuations ($\vel'$ and $\B'$), such that $\vel=\langle\vel\rangle+\vel'$ and $\B=\langle\B\rangle+\B'$ and consider again the evolution of large-scale magnetic fields and flows. Beyond its interaction with the mean flow $\langle\vel\rangle$, the evolution of the mean field $\langle\B\rangle$ now also depends on the fluctuation-induced electromotive force (e.m.f.) $\langle {\vel'} \times {\B}' \rangle$
 (where $\langle \cdot \rangle$ denotes a spatio-temporal or ergodic averaging
  process), while that of ${\vel}$ depends on the Reynolds and Lorentz stresses $\langle {\vel'} {\vel'} \rangle$ and $\langle {\B'} {\B'} \rangle$, respectively. Deep in the radiative zones, where small-scale fluctuations are very weak, these terms are likely not very large. However, in the vicinity of the interface between a radiative zone and a convective zone,  which is precisely the region of interest, these terms are expected to be much larger due to the ambient turbulence associated with convective overshooting motions. 
\par That being the case poses several problems. First, it is unlikely that Ferraro's isorotation theorem, which was derived under very restrictive conditions, continues to apply ``as is" for the large-scale field and flow in that region. As such, the magnetic field confinement, which many argued would be required to explain some of the observations, might no longer be necessary. Second, since the convection zone itself is likely the seat of a dynamo (which is one of the manifestations of the electromotive force) and therefore a {\it source} of magnetic field, confinement in the sense defined by \citet{CMG1993} or \citet{GM98} is not  possible in the first place. Instead, the magnetohydrodynamical coupling between the convective envelope and the radiative zone will likely be due to a combination of many different processes, including the previously discussed interaction of the large-scale flows and large-scale fields, but now also involving the interaction of the small-scale flows and small-scale fields, some of which are produced by dynamo action in the convection zone, and some of which are produced locally by the interaction of the overshooting motions with the large-scale radiative zone field.  
To study this complex problem, it is essential to break free of the traditional, two-dimensional, quasi-steady view of stellar magnetic fields, and to solve the full 3D, time-dependent MHD equations, including rotation, in a spherical coordinate system. The numerical complexity of this task is so formidable, that it is effectively presently unachievable. As such, we are forced to use simplified models, that may be limited spatially to a small region of the star, and/or ignore some of the aforementioned processes, and/or use governing parameters that are far from those of real stars. 
\par 
One of the most complete studies using this simplified approach was presented by \citet{Tobias2001} \citep[see also][]{Tobias98}.  Considering only a small region of the star located around the base of the convection zone, they performed 3D, nonlinear compressible simulations of penetrative convection at the interface between a convection zone and a radiative zone, and investigated its effect on the transport of magnetic fields.  For the wide range of parameters they examined, they found that any initial configuration of a large-scale horizontal magnetic field placed in the Cartesian box would be redistributed (or ``pumped") so that the majority of it resides in the radiative zone just below the overshoot zone, where it might eventually diffuse.  Note that \citet{Tobias2001} studied the effects of rotation on this turbulent confinement or pumping but did not attempt to include or study the effects of any large-scale meridional motions, which are central to the slow laminar confinement of \citet{GM98}.  
More recently, \citet{WB18} built on the work of \citet{Tobias2001} whereby they included a forced ``differential rotation" in the convection zone, which allowed them to drive and study the impact of large-scale meridional flows in the presence of turbulence. They found that, with careful parameter selection, the mean magnetic field in the radiation zone can be confined by the meridional flows, and therefore impose uniform rotation to that region. The turbulence in their simulations was however fairly mild and the overshooting motions on their own were not be able to confine the field in the absence of meridional flows, contrary to prior claims made by \citet{GR2007} using 2D numerical simulations, and by \citet{KitsRud08} using parametric models for magnetic pumping. 
Finally, \citet*{Strugarek} \citep[see also][]{BZ2006} performed fully 3D anelastic simulations  in a whole-Sun spherical geometry, which had the scope to allow for both slow laminar and fast turbulent confinement effects.   The choice of global geometry forced them to use model parameters that are too viscously dominated, however, and neither slow laminar nor fast turbulent confinement were found \citep[see the discussion of][]{AG13}.  
\par 
While all the case studies described above provide important insight into the problem, they are still largely preliminary, and cannot easily be used to make predictions on the magnetic coupling between the radiation zone and convection zones of stars other than the Sun. To do so would require a more fundamental understanding of the various processes involved, how they depend on stellar parameters, and how they interact with one another to affect the overall rotation profile of the star. By isolating some of these processes and studying them numerically in a systematic way, it is possible to gain a better understanding of their dependence on the model parameters  (which is crucial to the extrapolation from numerical simulations to stellar conditions). This is the approach we are taking in this series of papers.  We use 3D spherical Direct Numerical Simulations (DNS), and starting with the simplest problems, gradually add complexity.  Our initial work in \citet{Korre19}  omits rotation and examined the question of overshooting convection only.   In this work, we expand on \citet{Korre19} to study the interaction of the overshooting convection with an initially embedded magnetic field. We ignore the effects of rotation  in order to examine turbulent confinement in the absence of organized meridional flows (which, as discussed earlier, are an inevitable consequence of rotation). By doing so, we also ignore the effects of rotation on the convective eddies, which is perhaps not a good approximation near the base of the solar convection zone where the rotation rate $\Omega$ is comparable to the Brunt-V\"{a}is\"{a}l\"{a} frequency $N$. As such, our findings in this paper will not be directly applicable to the present-day Sun, but would be more relevant to more slowly rotating stars for which $\Omega << N$.  We also start by selecting convective parameters for which there is no dynamo, deferring the dynamo case to the next paper in the series. Finally we select an initial field whose amplitude is small enough not to affect the overshooting motions or the convection. As a result, the dominant interaction in this model is that of the overshooting turbulence on the initially large-scale, embedded field. 
\par This problem, even though very simple a priori, already reveals substantial complexity. We attempt to examine our results under the frameworks for magnetic transport that have been readily used before.  The most common of these is mean-field theory, where the electromotive force is modeled as $ \langle {\bf u'} \times {\B'} \rangle \simeq {\alpha} {\langle\B\rangle} + \boldsymbol{\gamma} \times {\langle\B\rangle}  + \beta \nabla \times {\langle\B\rangle}$, where ${\alpha}$ is the symmetric part of a tensor associated with the generation of large-scale poloidal magnetic fields from toroidal fields, $\boldsymbol{\gamma}$ is the anti-symmetric part of the same tensor related to magnetic pumping and $\beta$ is associated with turbulent diffusion. The $\beta$-effect \citep[see][]{KR80,CV91} is associated with the turbulent intensification of magnetic gradients and leads to a faster removal of the field. Some other well-known phenomenologies can be related to these ideas, such as the concept of ``magnetic flux expulsion"  \citep{Weiss66}, where magnetic flux is expelled from regions of circular streamlines and becomes concentrated at the edges of the convective eddies. The  ${\gamma}$- effect is the mean-field manifestation of ``diamagnetic pumping", which more generally refers to the transport of a magnetic field down a gradient of turbulent intensity, i.e. in the direction of decreasing turbulence. This process is particularly important in solar-type stars, where the turbulent diffusivity is much larger in the convection zone than in the radiative zone \citep[see, e.g.][]{KitsRud08}. 
\par The paper is organized as follows: In Section \ref{sec:setup}, we describe the model setup along with the initial conditions and the boundary conditions. In Section \ref{sec:results}, we present our numerical results  for four different Rayleigh numbers (where the Rayleigh number is the ratio of the buoyancy force over  the viscosity and the thermal diffusivity and measures the strength of the convective driving) and  compare
the DNS results with the exact solution of the induction equation in the absence of fluid motion, i.e. the purely diffusive case. In Section \ref{sec:mft}, we introduce a mean-field model approach to further explain and categorize the observed dynamics. Finally, in Section \ref{sec:discussion}, we
summarize and discuss our results, while in Section \ref{sec:conclusion}, we conclude with  their implications in the astrophysical context.

\section{Model set-up}
\label{sec:setup}

We are interested in studying the dynamics associated with the turbulent transport of magnetic fields between different regions on fast time-scales.  As discussed above, we omit rotation in order to isolate the turbulent confinement process from the laminar confinement associated with the slow rotationally-driven large-scale meridional flows. We solve the MHD version of the Spiegel-Veronis-Boussinesq equations \citep{SV} in a non-rotating spherical shell, knowing that even though the Boussinesq approximation might not be the most appropriate choice for astrophysical flows, it is still relevant when modeling deeper stellar interior regions, where the density stratification is small.
This allows us to explore a parameter regime with larger Rayleigh numbers and a lower Prandtl number and magnetic Prandtl number than those typically used in spherical compressible or anelastic simulations.
\par
Our  setup is similar to the one used in \citet{Korre19}, with a two-layered system that consists of a convectively unstable zone (CZ) lying on top of a stably stratified radiative zone (RZ).  The spherical shell has an outer radius $r_{\rm{o}}$, and  inner radius $r_{\rm{i}} = 0.2 r_{\rm{o}}$, while the CZ-RZ interface is located at $r_{\rm{t}} = 0.7 r_{\rm{o}}$. Within the radiative zone, we assume the presence of an initially compactly-contained  pre-existing dipole magnetic field (i.e. $\B_p = 0$ for $r > r_{\rm t}$)  and wish to study the evolution of this field.    We  solve the three-dimensional (3D) magnetohydrodynamic (MHD) Navier-Stokes equations under the   Boussinesq approximation, and assume that there is  a non-zero adiabatic background temperature gradient to account for  weak compressibility \citep{SV}. We use constant thermal expansion coefficient $\alpha$ (where $\alpha$ here is different from the one associated with the $\alpha-$ effect described  in Section \ref{sec:intro} and in Section \ref{sec:mft}), viscosity $\nu$, thermal diffusivity $\kappa$, adiabatic temperature gradient ${\rm{d}}T_{\rm{ad}}/dr$,   magnetic diffusivity $\eta$, and gravity $g$. Naturally, these quantities would  not be constant over the range $r = [0.2r_{\rm{o}},r_{\rm{o}}]$ in a star, but we make these assumptions for simplicity.  We use a fixed  flux inner boundary condition to mimic stellar conditions in which the flux is indeed set by the luminosity due to the nuclear burning in the stellar core, while  at the outer boundary we fix the temperature. Although this  is not a realistic outer boundary condition for solar-type stars (where complex radiative transfer processes would instead govern the thermal boundary conditions), we adopt these because they are simple, and expect  that they do not affect the convective dynamics in the bulk of the convective region. 
We  let $T(r,\theta,\phi,t)=T_{\rm{rad}}(r)+\Theta(r,\theta,\phi,t)$ where $T_{\rm{rad}}$ is the temperature profile our system would have under pure radiative equilibrium, and where $\Theta$ describes temperature fluctuations away from it. 
Under the Boussinesq approximation, there is a linear relationship between the temperature and density perturbations such that $\rho/\rho_{\rm{m}} = - \alpha \Theta$, where $\rho_{\rm{m}}$ is the mean density of the background fluid (again assumed constant). Then, the  governing MHD Boussinesq equations  are: 
\begin{eqnarray}
\nabla\cdot \vel=0,\\
\displaystyle\frac{\partial\vel}{\partial t}+\vel\cdot\nabla\vel=-\frac{1}{\rho_{\rm{m}}}\nabla p+\alpha \Theta g\boldsymbol{e_r}+\dfrac{1}{\rho_{\rm{m}}}\bold{j}\times\B+\nu\nabla^2\vel,\\
\nabla\cdot\B=0,\\
\displaystyle\frac{\partial\B}{\partial t}-\nabla\times(\vel\times\B)=\eta\nabla^2\B,\\
\displaystyle\frac{\partial\Theta}{\partial t}+\vel\cdot\nabla\Theta+u_r\left(\frac{{\rm{d}}T_{\rm{rad}}}{dr}-\frac{{\rm{d}}T_{\rm{ad}}}{dr}\right)=\kappa\nabla^2\Theta,
\end{eqnarray}
where $\vel=(u_r,u_{\theta},u_{\phi})$ is the velocity field, $\B=(B_r,B_{\theta},B_{\phi})$ is the magnetic field, $\bold{j}=({1}/{\mu_0})\nabla\times\B$ is the current density, $\mu_0$ is the vacuum permeability, and $p$ is the pressure  perturbation away from hydrostatic equilibrium.
As in \citet{Korre19}, because $\kappa$ is assumed to be constant,  we have to assume the existence of a heat source $H_{\rm s}(r)$ around $r_{\rm{t}}$ to set up the two-layered configuration whereby ${\rm{d}}T_{\rm{rad}}/dr - {\rm{d}}T_{\rm{ad}}/dr$ is negative in the CZ, and positive in the RZ. Then, in radiative equilibrium, we have
\begin{equation}
\label{eq:HS}
\kappa\nabla^2 T_{\rm{rad}}=-H_{\rm{s}}(r),
\end{equation}
and the background temperature gradient $T_{\rm{rad}}(r)$ is the solution of this equation, with the boundary conditions

\begin{equation}
-\kappa \frac{{\rm{d}}T_{\rm rad}}{dr} \bigg|_{r=r_{\rm{i}}}  = F_{\rm{rad}}, \quad T(r_{\rm{o}}) = T_{\rm{o}}, 
\end{equation} 
where $F_{\rm{rad}}$ is the temperature flux per unit area through the inner boundary. Integrating Equation (\ref{eq:HS}) once gives
\begin{equation}
\label{eq:HSb}
\displaystyle\kappa\frac{{\rm{d}}T_{\rm{rad}}}{dr}+\left(\dfrac{r_{\rm{i}}}{r}\right)^2F_{\rm{rad}}=-\dfrac{1}{r^2}\int_{r_{\rm{i}}}^r H_{\rm{s}}(r') r'^2 dr',
\end{equation}
and we therefore see that  we can essentially create any  chosen functional form for ${\rm{d}}T_{\rm{rad}}/dr$ with a suitable choice of $H_{\rm{s}}(r)$, without needing the exact expressions for $H_{\rm{s}}(r)$ and $T_{\rm{rad}}(r)$.

We non-dimensionalize the problem by using $[l]=r_{\rm{o}}$, $[t]=r_{\rm{o}}^2/\nu$, $[u]=\nu/r_{\rm{o}}$, $[B]=B_0$ and $[T]=|{\rm{d}}T_{\rm{o}}/dr-{\rm{d}}T_{\rm{ad}}/dr| r_{\rm{o}}$ as the unit length, time, velocity, magnetic field and temperature respectively, where ${{\rm{d}}T_{\rm{o}}}/{dr}\equiv{{\rm{d}}T_{\rm{rad}}}/{dr}|_{r=r_{\rm{o}}}$ is the radiative temperature gradient at the outer boundary and where $B_0$ sets the   amplitude of  the initial magnetic field (see Eq. (\ref{eq:Bp_IC})). With these units, we can write the non-dimensional equations as:
\begin{eqnarray}
{\nabla}\cdot{\vel}=0,\\
\displaystyle\frac{\partial{\vel}}{\partial{t}}+{\vel}\cdot{\nabla}{\vel}=-{\nabla}{ p}+\frac{\text{Ra}_{\rm{o}}}{\text{Pr}}{\Theta}\boldsymbol{e_r}+Q((\nabla\times\B)\times\B)+\nabla^2{\vel},\\
\nabla\cdot\B=0,\\
\label{eq:inducEq}
\displaystyle\frac{\partial\B}{\partial t}-\nabla\times(\vel\times\B)=\dfrac{1}{{\rm{Pm}}}\nabla^2\B,\\
\displaystyle\frac{\partial{\Theta}}{\partial{ t}}+{\vel}\cdot{\nabla}{\Theta}+\beta({r}){u_r}=\frac{1}{\rm{Pr}}{{\nabla}^2{\Theta}}.
\end{eqnarray}

\begin{figure}
\centering
\includegraphics[scale=0.38]{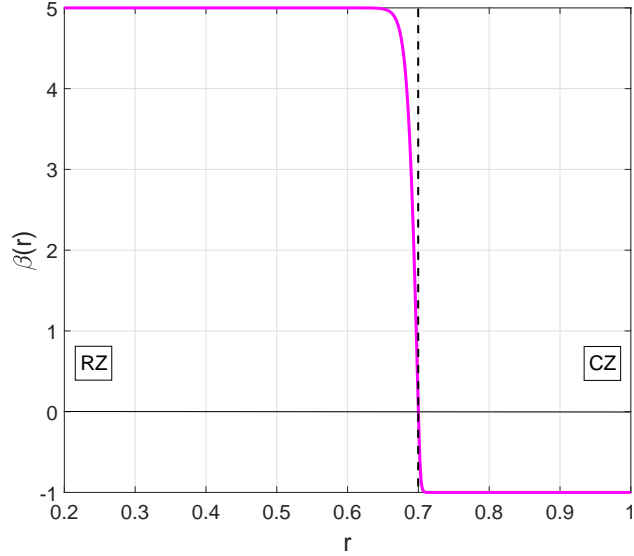}
\hspace{-13mm}
\caption{The profile of $\beta(r)$ versus the radius $r$, for $S=5$  and  $d_{\rm{out}}=0.003$.}
\label{fig:Figure_1}
\end{figure}

In all that follows, all the variables and parameters are now implicitly non-dimensional. This non-dimensionalization   introduces  the Prandtl number Pr and the global Rayleigh number Ra$_{\rm{o}}$ defined as
\begin{equation}
\text{Pr}=\displaystyle\frac{\nu}{\kappa}\quad \text{and}\quad \text{Ra}_{\rm{o}}=\displaystyle\frac{\alpha g\left|\displaystyle\frac{{\rm{d}}T_{\rm{o}}}{dr}-\frac{{\rm{d}}T_{\rm{ad}}}{dr}\right|r_{\rm{o}}^4}{\kappa\nu},
\label{eq:Rao}
\end{equation}
as well as the function $\beta(r)$ which is given by 

\begin{equation}
\beta(r)=\displaystyle\frac{\displaystyle\frac{{\rm{d}}T_{\rm{rad}}}{dr}-\displaystyle\frac{{\rm{d}}T_{\rm ad}}{dr}}{\displaystyle\left|\frac{d{T}_{\rm{o}}}{dr}-\displaystyle\frac{{\rm{d}}T_{\rm ad}}{dr}\right |}.
\end{equation}
By selecting a suitable  profile for $\beta(r) $ (implicitly selecting an appropriate  $H_{\rm{s}}(r)$ as described above) we can create a  convectively stable region for $r_{\rm{i}}\leq r<r_{\rm{t}}$ and an unstable region for $r_{\rm{t}}\leq r\leq r_{\rm{o}}$. Here, we choose the same prescription for the function $\beta(r)$ as in \citet{Korre19}, namely  
\begin{equation}
\label{eq:betaf}
\beta(r)=\left\{\begin{array}{l} 
\displaystyle -S\tanh\left(\frac{r-r_{\rm{t}}}{d_{\rm{in}}}\right) \mbox { when  }  r< r_{\rm{t}}, \\
\displaystyle -\tanh\left(\frac{r-r_{\rm{t}}}{d_{\rm{out}}}\right) \mbox { when  } r\geq r_{\rm{t}},
\end{array}\right.
\end{equation}
where $S$ is the stiffness parameter which measures the relative stability of the radiative zone to the convection zone and where $d_{\rm{in}}$ and $d_{\rm{out}}$ define the   transition  width between the two zones (see Figure \ref{fig:Figure_1}).  
Since  $\beta(r)$ and its derivative must be  continuous at $r_{\rm{t}}=0.7r_{\rm{o}}$, then  $d_{\rm{in}}=Sd_{\rm{out}}$. 
The function  $\beta$ can also be interpreted  as (minus) the ratio of the local Rayleigh number Ra$(r)$ to the global Rayleigh number Ra$_{\rm{o}}$, i.e.
\begin{equation}
\displaystyle\beta(r)=-\frac{\rm{Ra}(r)}{{\rm Ra}_{\rm{o}}},
\end{equation}
where 
\begin{equation}
\label{eq:betaRa}
\text{Ra}(r)=-\displaystyle\frac{\alpha g\left(\displaystyle\frac{{\rm{d}}T_{\rm{rad}}}{dr}-\frac{{\rm{d}}T_{\rm{ad}}}{dr}\right)r_{\rm{o}}^4}{\kappa\nu}.
\end{equation}
The minus sign in Equation (\ref{eq:betaRa}) ensures that Ra$(r)$ is positive in convective regions.  The presence of the magnetic field  introduces  the magnetic Prandtl number 
\begin{equation}
{\rm{Pm}}=\dfrac{\nu}{\eta}, 
\end{equation} 
and the Chandrasekhar number 
\begin{equation}
Q=\dfrac{B_0^2 r_{\rm{o}}^2}{\mu_0\rho_{\rm{m}}\nu^2},
\end{equation}
 which characterizes the relative importance of the Lorentz force to the viscous force.
In order to study the dynamics of the interaction of the overshooting motions with an initially contained dipolar magnetic field in the RZ, we have run 3D DNS solving the MHD Boussinesq equations in a spherical shell exactly as outlined above using the PARODY code \citep*{Parody2,Parody1}. 
 The chosen boundary conditions for the temperature  translate to   a no-flux boundary condition for the perturbations at the inner boundary,  $\partial \Theta/\partial r|_{r_{\rm{i}}}=0$, and  a zero temperature perturbation boundary condition  at the outer boundary, $\Theta(r_{\rm{o}})=0$. We employ stress-free boundary conditions for the velocity. For the magnetic field, we assume an electrically insulating outer boundary and a conducting inner core.
\par
\begin{figure}
\centering
\includegraphics[scale=0.4]{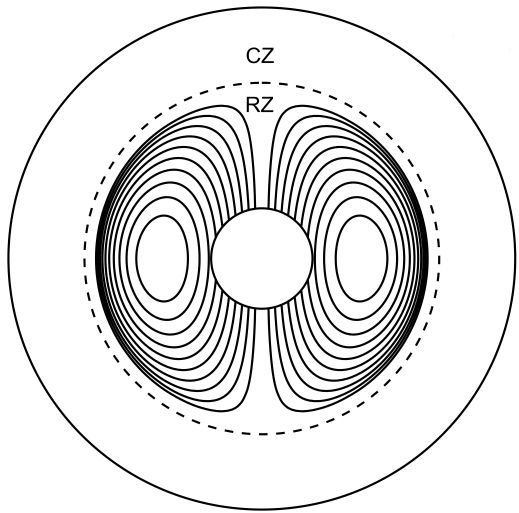}
\hspace{-13mm}
\caption{Initial configuration of the MHD simulation which starts with a dipole magnetic field  compactly contained in the stable radiative zone.}
\label{fig:Figure_2}
\end{figure}

  We initialize the MHD simulations from  the corresponding purely hydrodynamic  simulations  \citep[see][]{Korre19} that were evolved from a zero initial velocity and  small-amplitude temperature perturbations  until a statistically-stationary and thermally-relaxed state was achieved\footnote{We look both at the total kinetic energy per unit  volume in the domain, $E(t) = \frac{1}{2V}\int_V(u_r^2+u_{\theta}^2+u_{\phi}^2)dV$ (where $V$ is the volume of the spherical shell) and at the gradient of the  temperature perturbations at the surface $r_{\rm{o}}$ to check when that occurs.}. We then start the MHD simulation from the end of the thermally-equilibrated  hydrodynamic one, and    initialize it with  a purely poloidal dipole magnetic field initially contained in the stable zone below $0.65r_{\rm{o}}$ (Figure \ref{fig:Figure_2}) of the form: 
\begin{equation}
\label{eq:Bp_IC}
\B_p=B_0\nabla\times\nabla\times\left[\left(\dfrac{\sin(cr)}{(cr)^2}-\dfrac{\cos(cr)}{(cr)}\right)\sqrt{3}\cos\theta  \bold{\hat{r}}\right].
\end{equation}
The parameter $c$ is simply a geometric factor chosen to guarantee that $\B_p = 0$ at $r = 0.65 r_o$. To ensure this, $c=j_{(1,1)}/0.65r_o\approx 6.91$ (where $j_{(1,1)}$ is the first root of the function $j_1(r)=\sqrt{\pi/(2r)}J_{3/2}(r)$, where $J_{3/2}(r)$ is the Bessel function of order $3/2$).
\par All of the simulations reported here are for a fixed stiffness parameter $S=5$, a transition width $d_{\rm{out}}=0.003$\footnote{We note that in \citet{Korre19}, we presented a suite of  numerical simulations of overshooting convection (ignoring the effect of magnetism and rotation) where we varied both $S$ and $d_{\rm{out}}$ and studied the dependence of the overshooting dynamics on these input parameters.} , Pr$=0.1$ and Pm$=0.1$. We also fix  the Chandrasekhar number to be equal to $Q=0.01$, indicating that the initial magnetic field is relatively weak.  Note that the numerically achievable values of   the Prandtl number and the Rayleigh number are not astrophysically realistic (e.g. for the Sun:  Pr$_{\odot}\sim 10^{-6}$ in the solar tachocline, and  Ra$_{\odot}\sim 10^{23}$) due to the computational constraints arising from the required spatial and temporal resolution of the simulations. However, the magnetic Prandtl number Pm is of the order of the solar value, which is approximately equal to Pm$=0.1$ at the bottom of the solar CZ. 
\par With these chosen parameters, there is no dynamo action, hence the field can only decay. In this paper, we  focus solely on studying  how the overshooting turbulent motions can affect the transport and overall evolution of a weak magnetic field. We vary Ra$_{\rm{o}}$ over three orders of magnitude (Ra$_{\rm{o}} = 10^6$, $10^7$,   $10^8$, and $10^9$), and study the time-dependent evolution of the magnetic field as each simulation proceeds.

\begin{table*}

\begin{tabular}{lccccccc}
  &Ra$_{\rm{o}}$  & $N_r$ &$N_{\theta}$ &$N_{\phi}$ &  $\delta_{\rm u}$
 &  $\lambda$ & Rm\\ 
\hline
\\
&   $10^6$  &300 & 192 & 192 & 0.13  &145 & 8
\\ \\
&    $10^7$  &400 & 288 & 320 & 0.094 &191 & 22
 \\ \\
&    $10^8$ &585 & 516 & 640  & 0.069&243 & 50
\\ \\
&   $10^9$ &585 & 516 & 640 & 0.049 &319 & 112 \\ 
\hline
\end{tabular}
\caption{\label{tab:MHDdata} Table with  input and output parameters and resolution for $S=5$, $d_{\rm{out}}=0.003$,  Pr$=0.1$, and Pm$=0.1$. The resolution is provided in number of  meshpoints $N_r$, $N_{\theta}$ and $N_{\phi}$,  $\delta_{\rm u}$ is the overshoot length-scale (for more details see Section \ref{sec:results}), $\lambda$ is the  measured decay rate of the amplitude of the dipole field in the exponential decay phase (see Section \ref{sec:mft}), and Rm is the magnetic Reynolds number (see Section \ref{sec:discussion}).}
\end{table*}

\section{Numerical results}
\label{sec:results}
 
\begin{figure}
\centering
\includegraphics[scale=0.35]{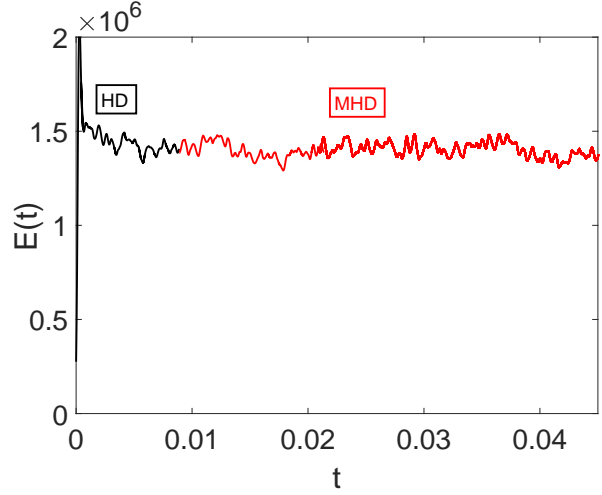}
\caption{\label{fig:Figure_3} Total kinetic energy per unit volume $E(t)$ versus time for the run with Ra$_{\rm{o}}=10^8$. The black color corresponds to the hydrodynamic simulation (HD) and the red color corresponds to the MHD part of the simulation (MHD), i.e. after we have added the magnetic field.}
\end{figure}

\begin{figure}
\centering
\includegraphics[scale=0.5]{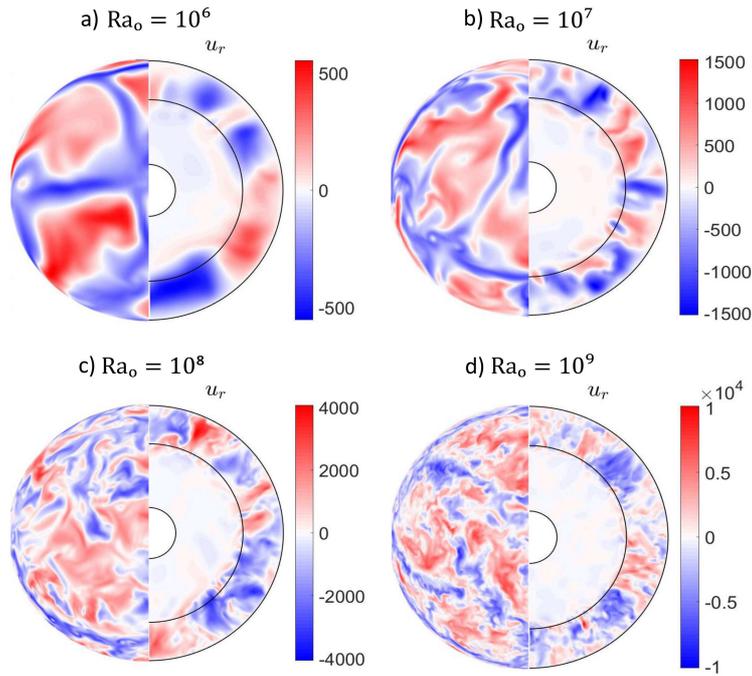}
\caption{\label{fig:Figure_4} {Snapshots of $u_r$. In each panel, the left part shows the $u_r$
field close to the outer radius, while the right
part shows the same field $u_r$  at a selected longitude at a) Ra$_{\rm o}=10^6$, b) Ra$_{\rm o}=10^7$, c) Ra$_{\rm o}=10^8$, and d) Ra$_{\rm o}=10^9$. The inner black line represents the base of the convective region at $r_{\rm t}$.}}
\end{figure}

 We begin by  exploring the dynamics observed in a typical simulation with  Ra$_{\rm{o}}=10^8$.
 In Figure \ref{fig:Figure_3}, we present the total kinetic energy per unit volume $E(t)$ against time $t$. The black line corresponds to $E$ from the purely  hydrodynamic run (HD) from which the MHD simulation was restarted, while the red line is from the same simulation after adding the magnetic field (MHD). The kinetic energy  does not change noticeably  between the HD run and the MHD run, indicating that the inclusion of the field does not affect the convective dynamics significantly. This is expected since the field is initially weak, and  decays with time.  In Figure \ref{fig:Figure_4}, we present snapshots  of  the radial velocity $u_r$. In each panel,
the left hemisphere shows the velocity field on a spherical
shell close to the upper boundary at $r\approx 0.89r_{\rm o}$, illustrating the convective motions near the surface. The right hemisphere is a meridional
slice showing the radial velocity for a selected longitude as a function of $r$ and $\theta$ at a) Ra$_{\rm o}=10^6$, b) Ra$_{\rm o}=10^7$, c) Ra$_{\rm o}=10^8$ and d) Ra$_{\rm o}=10^9$ taken during the statistically stationary state. In all cases,  we notice  that the degree of turbulence in the CZ increases with increasing Ra$_{\rm o}$ with higher Rayleigh numbers resulting in stronger eddies with a wider range of scales. Also, in the right hemispheres, we see that the convective motions generated within the convective region overshoot some distance beyond the bottom of the CZ (represented by the inner black line).
 \par
\begin{figure*}
\centering
\includegraphics[scale=0.8]{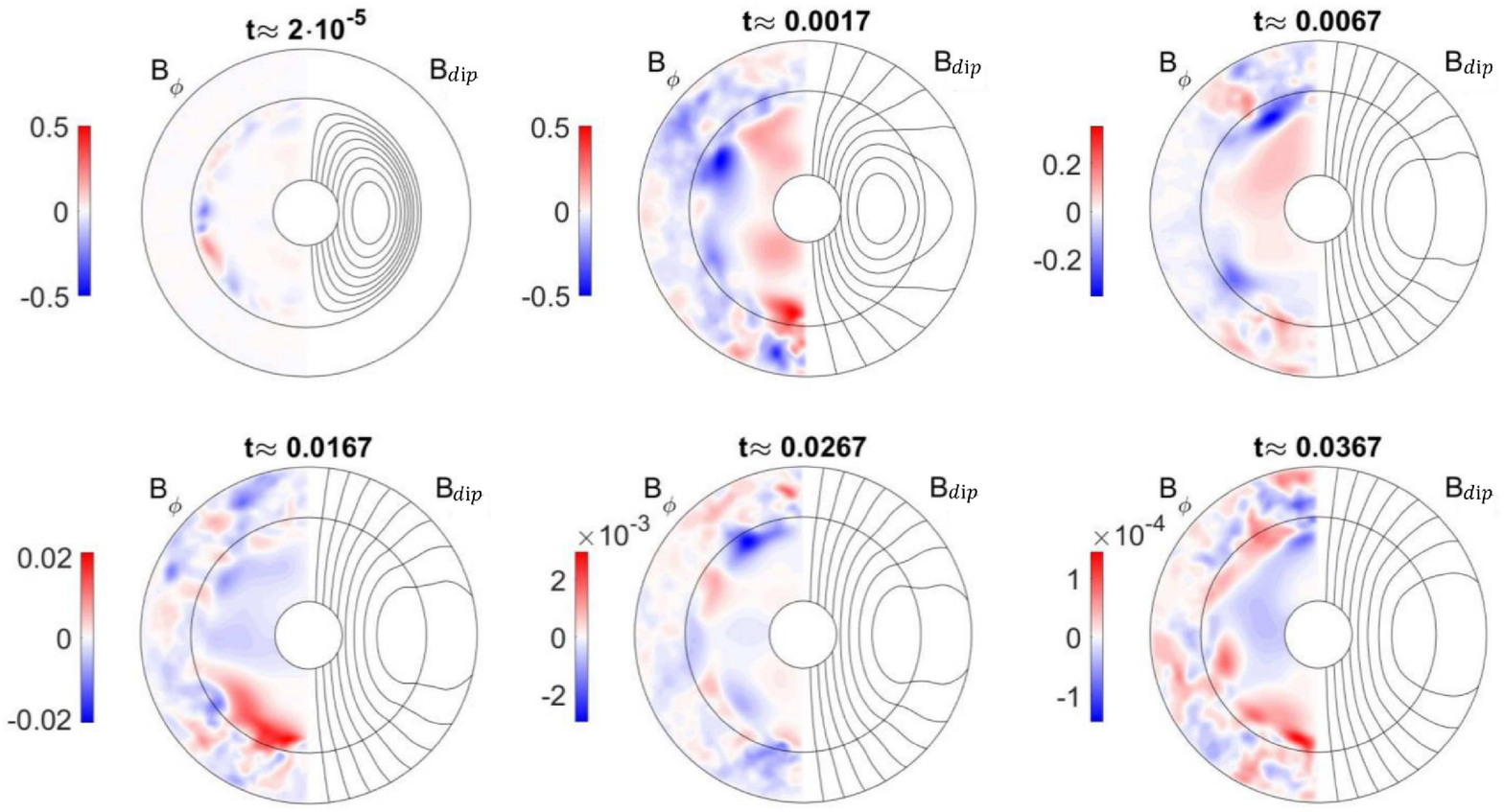}
\caption{\label{fig:Figure_5}Contours of the axisymmetric $B_{\phi}$ and the dipole component $B_{\rm dip}$ for six different times of the run with Ra$_{\rm{o}}=10^8$.}
\end{figure*}

  To illustrate   the evolving geometry of the large-scale  magnetic field, we plot contours of the  dipole component of the poloidal  field   along with the azimuthally-averaged  toroidal component of the magnetic field $B_{\phi}$  at Ra$_{\rm{o}}=10^8$ at different representative times (Fig. 
 \ref{fig:Figure_5}). The first panel shows the solution very close to the initial condition, where most of  the dipole is still contained compactly within $r<0.7r_{\rm o}$. At $t\approx 0.0017$, the field has already diffused outwards a little, come into contact with the overshooting motions, and started ``opening up" into the CZ. By $t\approx 0.0067$, all of the field lines have now  opened up into the  CZ. This  leads to the appearance of an unconfined configuration, i.e. a  state  where the  dipole field lines have infiltrated the CZ substantially leaving very few (if any) closed field lines in the RZ. 
 Note how the overall field geometry then barely changes after this point, suggesting that it may have settled into a particular eigenmode of the induction equation. The field amplitude is still decaying with time, however.  At $t\approx 0.00002$,  $B_{\phi}$    is of order unity, but decreases by three orders of magnitude by the time $t\approx 0.0367$. 

\par We now analyze more quantitatively the results of the simulations, in order to get a better understanding of the spatio-temporal evolution of the field.   In what follows, it will be informative  to compare our numerical  results to a hypothetical case without convection, in which the initial field merely diffuses away. In the absence of any fluid motion, the initial magnetic field evolves with time according to 
\begin{equation}
\label{eq:Bdiff}
\dfrac{\partial \B_{\rm diff}}{\partial t}=\dfrac{1}{\rm{Pm}}\nabla^2\B_{\rm diff},
\end{equation}  
with Pm$=0.1$.
 We can solve this equation semi-analytically, to compute the evolution of the magnetic energy  in the absence of convection  (see Appendix \ref{app:AppA}).
In Figure \ref{fig:Figure_6}, we compare the volume average of the  magnetic  energy  of the purely diffusive case along with the same quantity computed  in the fully nonlinear convective simulation. Note that the dipole component of the magnetic field is given by
\begin{equation}
\label{eq:dipole}
\B_{\rm dip}(r,\theta)=(B_{r_{\rm dip}},B_{\theta_{\rm dip}})=\left(-\dfrac{1}{r}L_2A,\quad \dfrac{\partial}{\partial\theta}\left(\dfrac{1}{r}\dfrac{\partial}{\partial r}(rA)\right)\right),
\end{equation}
where  \begin{equation}
\label{eq:L2}
L_2=\frac{1}{\sin \theta} \frac{\partial}{\partial \theta} \left( \sin \theta \frac{\partial}{\partial \theta} \right),
\end{equation}
 and $A(r,\theta,\phi)=y_{B_{\rm dip}}(r)Y_1^0(\theta,\phi)$, where $y_{B_{\rm dip}}(r)$ is the amplitude of the dipole evolved in the code and  $Y_1^0(\theta,\phi)=\sqrt{3}\cos\theta$ is the spherical harmonic with degree $(l,m)=(1,0)$ corresponding to that dipole mode. Then, we define the spherically-averaged  dipole magnetic energy in the simulation as
\begin{equation}
\label{eq:Edipr}
\bar{E}_{\rm dip}(r)=\dfrac{Q}{2}\int_0^{\pi}\left(B_{r_{\rm dip}}^2+B_{\theta_{\rm dip}}^2\right)\sin\theta d\theta
\end{equation}
such that the volume-averaged dipole magnetic energy in the spherical shell is 
\begin{equation}
\label{eq:Edipt}
E_{\rm dip}=\dfrac{\int_{r_{\rm{i}}}^{r_{\rm{o}}} \bar{E}_{\rm dip}r^2 dr}{{\int_{r_{\rm{i}}}^{r_{\rm{o}}}r^2dr}}.
\end{equation}
In Figure \ref{fig:Figure_6}, we see that the evolution of the magnetic energy in the dipole field in the simulation $E_{\rm dip}$ coincides more or less with that of the purely diffusive solution $E_{\rm diff}$ (defined similarly to $E_{\rm dip}$, but with $\B_{\rm dip}$ replaced by $\B_{\rm diff}$) for $t\leq 0.01$, but as time evolves beyond this point, $E_{\rm dip}$ begins to  decrease much faster than  $E_{\rm diff}$. Notably,  the decrease is faster for higher values of Ra$_{\rm{o}}$. Furthermore, we notice that after some  adjustment period, $E_{\rm dip}$ decays exponentially with a well-defined, constant decay rate (that depends on Ra$_{\rm o}$). This confirms our conclusions from the visual inspection of Figure \ref{fig:Figure_5}, and demonstrates  that the dipole field in each simulation eventually settles into an eigenmode of the induction equation. This is consistent with the fact that the field is weak (so the Lorentz force in the momentum equation is negligible), so the induction equation is essentially kinematic (linear in $\B$).

\begin{figure}
\centering
\includegraphics[scale=0.3]{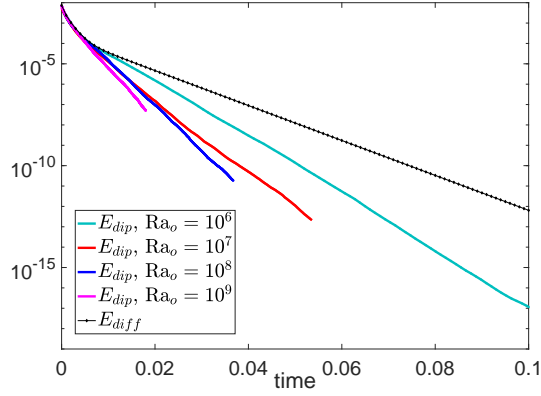}
\caption{\label{fig:Figure_6}Non-dimensional magnetic energy of the dipole, ${E}_{\rm dip}(t)$ for Ra$_{\rm{o}}=10^6$, $10^7$, $10^8$ and $10^9$ plotted along with $E_{\rm diff}(t)$ against the time $t$.}
\end{figure}

One could rightfully ask whether this significant and rapid loss of energy  in the dipole (compared with the purely diffusive case) could  potentially be attributed to  Tayler instabilities. Past studies  \citep[e.g.][]{MTinstab73,Wright73,Spruit99,Braithwaite} have shown that a purely  poloidal  field with closed field lines within a stable radiative zone is unstable to non-axisymmetric perturbations, and these MHD instabilities can lead to a substantial reduction of  the magnetic  energy. However, we do not observe any such instability in this work, because our initial field is too weak. Indeed, Tayler instabilities grow on an Alfv\'{e}nic time-scale,  which is $t_A=1/\sqrt{Q}=10$ for the parameter $Q$ selected in this set of simulations.  
  This is clearly longer   than both the magnetic diffusion time-scale and the thermal diffusion time-scale in our system, which are $t_{\eta}=0.1$ and $t_{\kappa}=0.1$ respectively. 
This, and the fact that the decay rate depends on Ra$_{\rm{o}}$, lead us to the conclusion that the faster-than-diffusive decay is a consequence of the convective motions acting on the field, rather than other types of instabilities within the radiative zone. 
\par
\begin{figure*}
\centering
\includegraphics[scale=0.4]{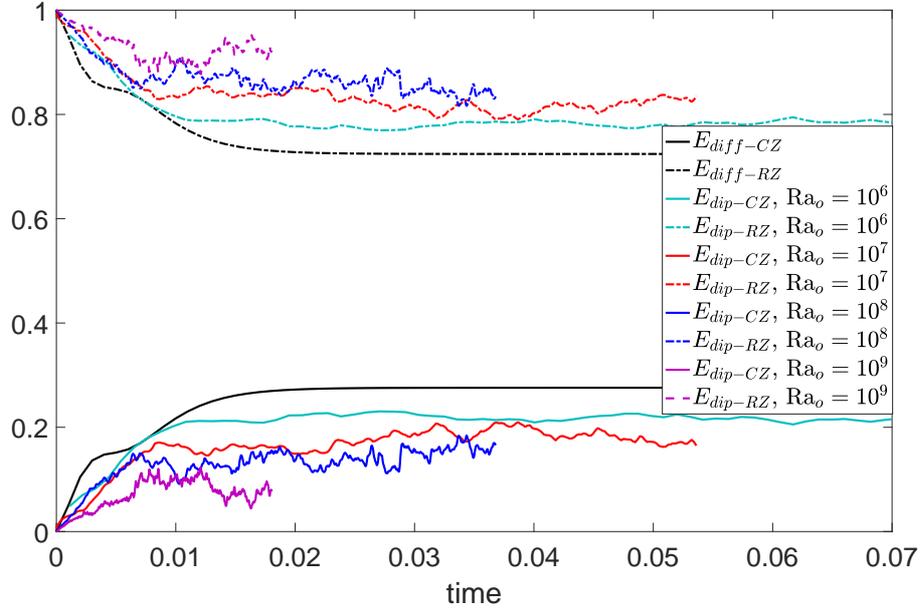}
\caption{\label{fig:Figure_7}Plot of $E_{\rm dip-RZ}$ and $E_{\rm dip-CZ}$ for the runs with Ra$_{\rm{o}}=10^6$, Ra$_{\rm{o}}=10^7$, Ra$_{\rm{o}}=10^8$ and Ra$_{\rm{o}}=10^9$ along with the purely diffusive $E_{\rm diff-RZ}$ and $E_{\rm diff-CZ}$ with Pm$=0.1$.}
\end{figure*}

 Another quantitative way of examining the evolution of the initial field  is to look at the radial distribution of the magnetic energy over time between the CZ and the RZ  and compare it with the purely diffusive case. We define the fractional magnetic energy of the dipole in the RZ, $E_{\rm dip-RZ}$, and  in the CZ,  $E_{\rm dip-CZ}$, respectively, as

\begin{equation}
\label{eq:EdipRZ}
\displaystyle E_{\rm dip-RZ}=\dfrac{\displaystyle\int_{r_{\rm{i}}}^{r_{\rm{t}}}\bar{E}_{\rm dip}r^2dr}{\displaystyle\int_{r_{\rm{i}}}^{r_{\rm{o}}}\bar{E}_{\rm dip}r^2dr},
\end{equation}
and
\begin{equation}
\label{eq:EdipCZ}
\displaystyle E_{\rm dip-CZ}=\dfrac{\displaystyle\int_{r_{\rm{t}}}^{r_o}\bar{E}_{\rm dip}r^2dr}{\displaystyle\int_{r_{\rm{i}}}^{r_{\rm{o}}}\bar{E}_{\rm dip}r^2dr}.
\end{equation}
In Figure \ref{fig:Figure_7}, we plot  $E_{\rm dip-RZ}$ and  $E_{\rm dip-CZ}$  versus time,  for the runs with Ra$_{\rm{o}}=10^6,10^7,10^8,10^9$ as well as for the purely diffusive case, where we calculate $E_{\rm diff-CZ}$ and $E_{\rm diff-RZ}$ in a similar way. 
At $t=0$, the magnetic field is fully confined in the RZ, so $E_{\rm dip-CZ} = 0$ and $E_{\rm dip-RZ} = 1$. As $t$ increases, we see that in all cases (both purely diffusive and the numerical simulations at increasing Ra$_{\rm{o}}$), there is an initial decrease in $E_{\rm dip-RZ}$ and a concurrent increase in $E_{\rm dip-CZ}$, which corresponds to the initial stages of the evolution where the  dipole field begins to diffuse into the convection zone. Around $t=0.015$, the purely diffusive case starts relaxing towards the slowest-decaying  radial eigenmode of the diffusion equation, and eventually $E_{\rm diff-RZ}$ and $E_{\rm diff-CZ}$ asymptote  to two constants with $E_{\rm diff-RZ}>E_{\rm diff-CZ}$, indicating that the magnetic energy distribution is then  decreasing self-similarly. The same  general stages of evolution are seen in the DNS.   The fractional  energies of  the full 3D calculations also  asymptote to two statistically-stationary constants at each Ra$_{\rm{o}}$, which confirms that the dipole field in each simulation  has  relaxed to the slowest decaying eigenmode of the induction equation. This is consistent with our findings that the dipole energy is decreasing exponentially in Figure \ref{fig:Figure_6}. Crucially, however,  we  find that the ratio $E_{\rm dip-RZ}/E_{\rm dip-CZ}$ increases substantially with Ra$_o$, which could be interpreted as the field being increasingly more contained and therefore ``confined" in the RZ than in the diffusive case.
\par
\begin{figure}
\centering
\includegraphics[scale=0.35]{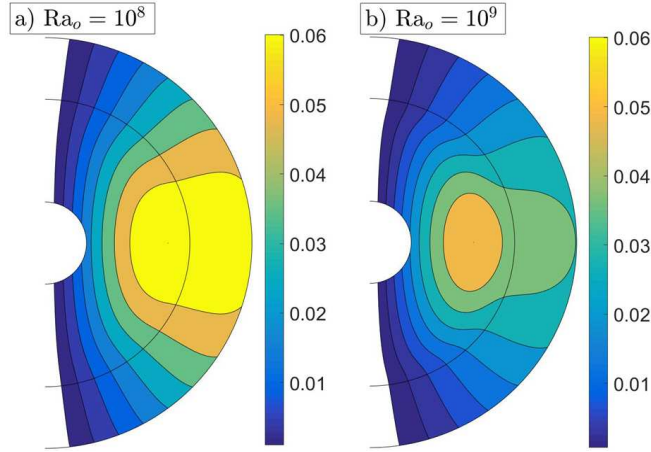}
\caption{\label{fig:Figure_8} Contour plot of the dipole field lines in the exponential decay phase (see text for more details) for a) Ra$_{\rm{o}}=10^8$, and b) Ra$_{\rm{o}}=10^9$. }
\end{figure}

In an effort to understand this behaviour,  we now focus on the    Ra$_{\rm{o}}=10^9$ case  for which  $E_{\rm dip-RZ}/E_{\rm dip-CZ}$  is the largest and  begin by comparing the properties of its dipole eigenmode to that of the Ra$_{\rm{o}} = 10^8$ case. We extract this eigenmode by computing the weighted time-average  $\bar{\B}_{\rm dip}$ as
\begin{equation}
\label{eq:Bdipnorm}
\bar{\B}_{\rm dip}(r,\theta)=\dfrac{1}{N}\sum_{t=1}^N\left(\dfrac{\B_{\rm dip}(r,\theta)}{\B_{r_{\rm dip}}(r_{\rm i},0)}\right)_t,
\end{equation} 
where $N$ is the number of  available snapshots of the DNS at different  times  after $t\approx 0.014$  (i.e. in the exponential decay phase), and  where ${\B}_{\rm dip}(r_{\rm i},0)$ is the amplitude of the radial component of the dipole field at the inner boundary ($r = r_{\rm i}$) at the pole, used here to normalize the decaying eigenmode. Figures  \ref{fig:Figure_8}a and \ref{fig:Figure_8}b show selected field lines of $\bar{\B}_{\rm dip}$ for the Ra$_{\rm{o}} = 10^8$ and Ra$_{\rm{o}} = 10^9$ cases, respectively. In both cases,  we  see that the field lines have clearly  diffused into the CZ. However, while all field lines are clearly open in the Ra$_{\rm{o}} = 10^8$ case,   some of the field lines  appear to remain closed in the RZ in the Ra$_{\rm{o}} = 10^9$ case, showing further evidence for a more confined state at higher Ra$_{\rm{o}}$ (in an average sense).  

\par
\begin{figure*}
\centering
\includegraphics[scale=0.55]{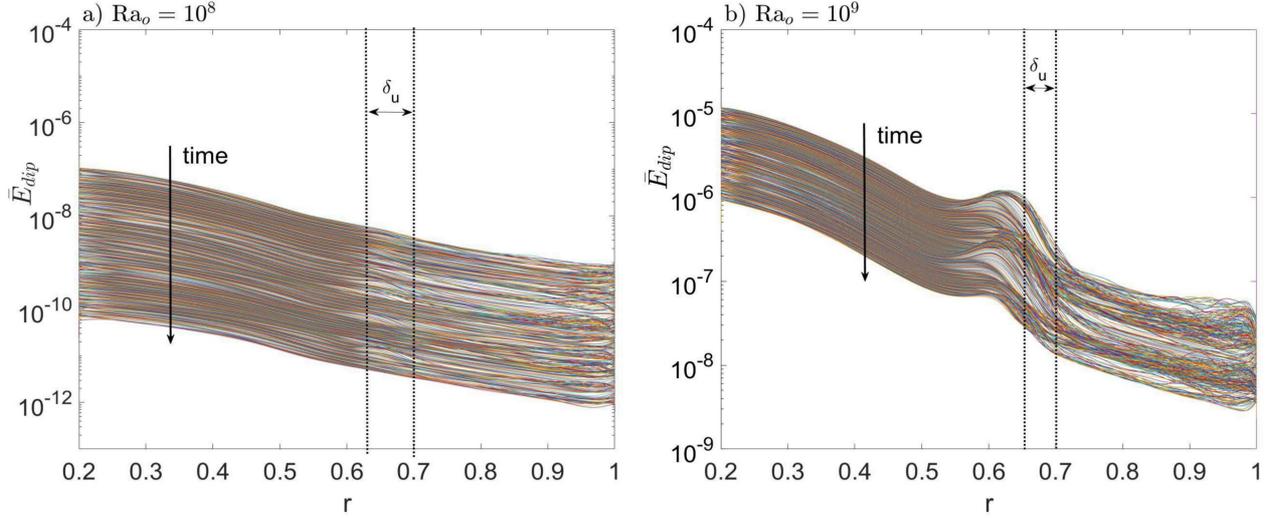}
\caption{\label{fig:Figure_9}  Plot of the  spherically-averaged energy of the dipole mode  $\bar{E}_{\rm dip}$ versus $r$ at different times in the exponential decay phase for a) Ra$_{\rm{o}}=10^8$ and  b) Ra$_{\rm{o}}=10^9$.}
\end{figure*}

 Figure \ref{fig:Figure_9}a and   Figure \ref{fig:Figure_9}b  compare more quantitatively the evolution of the dipole fields in the Ra$_{\rm{o}}= 10^8$ and $10^9$ simulations, by  showing $\bar{E}_{\rm dip}$   as a function of radius at different times after the transient state.  We see that for the  Ra$_{\rm{o}}=10^8$ case, $\bar{E}_{\rm dip}$ monotonically decreases from the RZ  outward and decays with time. For the Ra$_o=10^9$ run, by contrast, we observe a ``bump"  representing an excess of dipole energy that can be shown to be located just below the  convective overshoot region. Indeed, as shown by \citet{Korre19}, it is possible to characterize the overshoot depth $\delta_{\rm u}$ using the stopping distance of the strongest downflows originating from the CZ. The accumulation of the dipole energy  in Fig. \ref{fig:Figure_9}b appears to reside just below the radius $r_{\rm t} - \delta_{\rm u}$.  This suggests that it is likely associated with the ejection of the field from the CZ by the stronger  downflows, akin to  earlier ideas of ``magnetic pumping"  \citep[e.g.][]{Tobias98,Tobias2001}.  This effect does not appear to be significant at lower Ra$_{\rm{o}}$ and only appears here at Ra$_{\rm{o}}= 10^9$, supporting the notion that there must be a  substantial change in the interaction between the magnetic field and the convective motions  at the highest Ra$_{\rm{o}}=10^9$ case.

\section{A mean-field model for the dynamics}
\label{sec:mft}

\par The two effects found above -- the increasingly rapid decay of the dipole field, and the increasing fraction of dipole magnetic energy in the RZ with increasing Ra$_{\rm{o}}$  -- are inevitably due to  induction effects.     Much work has been done to characterize and classify the behaviour of the e.m.f. term  $\vel\times\B$  using a mean-field approach, as described in the introduction.   We have found that a mean-field model that includes both a magnetic pumping term (typically called a $\gamma-$ effect) and a turbulent diffusion term (typically called a $\beta-$ effect) appears to be sufficient to correctly capture the effects of the turbulent flow on the mean field. We therefore now attempt to discern the effect of each one of these processes on the dipole field and their dependence on Ra$_{\rm{o}}$. 
\par The mean field induction equation for the large-scale field $\langle\B\rangle$ (where $\langle\cdot\rangle$ is  an azimuthal  average to distinguish between large scales of interest and small-scale turbulent motions), with these terms, is given by
\begin{equation}
\label{eq:MF}
\dfrac{\partial \langle\B\rangle}{\partial t}=\nabla\times({\alpha}\langle\B\rangle+\boldsymbol{\gamma}\times\langle\B\rangle-\eta_{\rm T}\nabla\times\langle\B\rangle)+\eta\nabla^2\langle\B\rangle,
\end{equation}  
where ${\alpha}\langle\B\rangle$ is  the well-known   ``$\alpha-$ effect"  that is derived from the symmetric part of the mean-field tensor and allows the regeneration of large-scale poloidal magnetic fields, $\boldsymbol{\gamma}\times\langle\B\rangle$ is derived from the  antisymmetric  part of the mean-field tensor and is related to the concept of magnetic pumping since $\boldsymbol{\gamma}$ looks like a velocity,   and $\eta_{\rm T}\nabla\times\langle\B\rangle$ is associated with turbulent diffusion, where   $\eta_{\rm T}$ (often termed $\beta$) is the turbulent diffusivity. Even though the convection is not isotropic, we make the further approximation that $\boldsymbol{\gamma}=-\dfrac{1}{2}\nabla\eta_{\rm T}$, derived for 3D  nearly isotropic turbulence  \citep{KR80, KR92, KitsRud08}. 
\par Although Eq. (\ref{eq:MF}) is usually written for any ``mean-field", in what follows we take $\langle \B \rangle$ to be $\B_{\rm dip}$. Then, as before, we define $\A$  to be the potential field such that $\B_{\rm dip}=\nabla\times\A$, with
\begin{equation}
\label{eq:Afield}
\A=(0,0,-\partial A/\partial\theta)=(0,0,\sqrt{3}y_{B_{\rm dip}}(r)\sin\theta),
\end{equation}
and thus obtain an evolution equation for $\A$, 

\begin{equation}
\label{eq:potA}
\dfrac{\partial(\nabla\times\A)}{\partial t}=\nabla\times[{\alpha}(\nabla\times\A)+\boldsymbol{\gamma}\times(\nabla\times\A)-\eta_{\rm T}\nabla\times(\nabla\times\A)]+\eta\nabla^2(\nabla\times\A).
\end{equation} 
Projecting Eq. (\ref{eq:potA}) onto the $\phi$ direction while also assuming that $\eta_{\rm T}=\eta_{\rm T}(r)$, i.e. that the turbulent diffusivity only has a radial dependence, and 
substituting the ansatz (\ref{eq:Afield}) then yields
\begin{equation}
\label{eq:MFeqdipole1}
\sqrt{3}\sin\theta\dfrac{\partial y_{B_{\rm dip}}}{\partial t}=
-\dfrac{1}{2}\dfrac{d\eta_{\rm eff}}{dr}\dfrac{\partial}{\partial\theta}\left(\dfrac{1}{r}\dfrac{\partial}{\partial r}(r y_{B_{\rm dip}}\sqrt{3}\cos\theta)\right)+\eta_{\rm eff}\left(\nabla^2(y_{B_{\rm dip}}\sqrt{3}\sin\theta)-\dfrac{y_{B_{\rm dip}}\sqrt{3}\sin\theta}{r^2\sin^2\theta}\right),
\end{equation}
where $\eta_{\rm eff}$ is the effective diffusivity (which is the sum of the microscopic diffusivity equal to $1/$Pm and the turbulent diffusivity).
Once the simulation has reached the exponentially decaying eigenstate discussed earlier, we can assume that ${\partial y_{B_{\rm dip}}}/{\partial t}=-\lambda y_{B_{\rm dip}}$, where $\lambda$ is the measured decay rate of the amplitude of the dipole  in the exponential decay phase (see Table \ref{tab:MHDdata}). Then, with a few algebraic manipulations we obtain 
\begin{equation}
\label{eq:tranlambda}
-\lambda y_{B_{\rm dip}}=
\dfrac{1}{2}\dfrac{d\eta_{\rm eff}}{dr}\dfrac{1}{r}\dfrac{\partial}{\partial r}(r y_{B_{\rm dip}})+\eta_{\rm eff}\left(\dfrac{\partial^2y_{B_{\rm dip}}}{\partial r^2}+\dfrac{2}{r}\dfrac{\partial y_{B_{\rm dip}}}{\partial r}-\dfrac{2}{r^2}y_{B_{\rm dip}}\right).
\end{equation}
This equation can be used to infer the quantity $\eta_{\rm eff}(r)$, given $y_{B_{\rm dip}}$ and $\lambda$ extracted from the DNS. However, the profiles of $y_{B_{\rm dip}}(r,t)$  at individual time-steps are too noisy, and so cannot be used ``as is". We therefore first perform a weighted time-average of $y_{B_{\rm dip}}$ in the exponentially decaying phase, as 

\begin{equation}
\label{eq:ydip}
\bar{y}_{B_{\rm dip}}(r)=\dfrac{1}{N}\sum_{t=1}^{N}\left(\dfrac{y_{B_{\rm dip}}(r,t)}{y_{B_{\rm dip-av}}(t)}\right),
\end{equation}
 where
\begin{equation}
\label{eq:ydipav}
y_{B_{\rm dip-av}}(t)=\left|\dfrac{\int_{r_{\rm{i}}}^{r_{\rm{o}}}y_{B_{\rm dip}}(r,t)r^2dr}{\int_{r_{\rm{i}}}^{r_{\rm{o}}}r^2dr}\right|.
\end{equation}
This effectively extracts the eigenmode of the problem, as we did in Eq. (\ref{eq:Bdipnorm}) for $\B_{\rm dip}$. Now,  let $f(r)=r \bar{y}_{B_{\rm dip}}(r)$
 Then Eq. (\ref{eq:tranlambda}) becomes
\begin{equation}
\label{eq:lambdaf}
-\lambda f=\dfrac{1}{2}\dfrac{d\eta_{\rm eff}}{dr}\dfrac{df}{dr}+\eta_{\rm eff}\left(\dfrac{d^2f}{dr^2}-\dfrac{2f}{r^2}\right),
\end{equation}
which is a first-order ordinary differential equation for $\eta_{\rm eff}(r)$, given $f(r)$.
Note that the problem is singular at the point $r = r_f$ where  $df/dr=0$. However, the equation does have a regular solution, which we compute by  enforcing the internal boundary condition 
\begin{equation}
\label{eq:BC}
\eta_{\rm eff}(r_f)=\dfrac{\lambda f(r_f)}{\dfrac{d^2f(r_f)}{dr^2}-\dfrac{2 f(r_f)}{r_f^2}}
\end{equation}
at that point, and  numerically integrating Eq. (\ref{eq:lambdaf}) inward for $r\le r_f$ and outward for $r\ge r_f$. Profiles of $\eta_{\rm eff}(r)$ obtained in this manner are shown and discussed below.

\par
\begin{figure}
\centering
\includegraphics[scale=0.32]{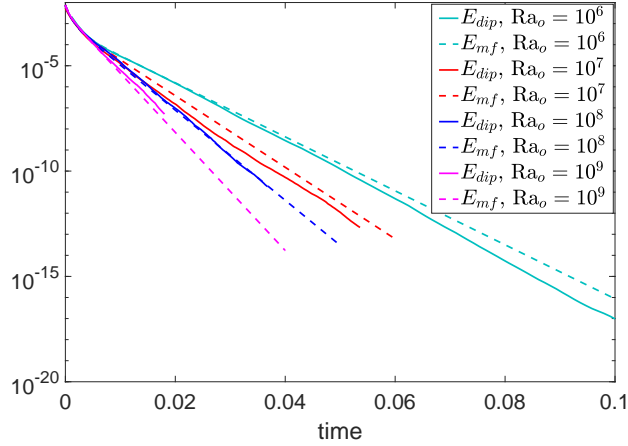}
\caption{\label{fig:Figure_10}  Profile of the dipole magnetic energy against time comparing  $E_{\rm dip}$ from the simulation data for the four values of Ra$_{\rm{o}}$ (solid lines)  with $E_{\rm mf}$ calculated using Eq. (\ref{eq:Bdiffeta}) (dashed lines).}
\end{figure}

 
In order to cross-check the validity of the procedure, after computing $\eta_{\rm eff}$ we also   solve the forward mean-field problem 

\begin{equation}
\label{eq:Bdiffeta}
\dfrac{\partial \langle\B_{\rm mf}\rangle}{\partial t}=\nabla\times\left(-\dfrac{1}{2}\nabla\eta_{\rm T}\times\langle\B_{\rm mf}\rangle-\eta_{\rm T}\nabla\times\langle\B_{\rm mf}\rangle\right)+\eta\nabla^2\langle\B_{\rm mf}\rangle,
\end{equation}
subject to the same initial  and boundary conditions on the field as in the DNS.  
In Figure \ref{fig:Figure_10}, we  plot the volume average of the dipole magnetic energy $E_{\rm dip}$ against the time  for the four  Ra$_{\rm{o}}=10^6, 10^7,10^8, 10^9$ cases from our DNS  (solid lines) along with the dipole magnetic energy $E_{\rm mf}$ of the mean-field forward problem  computed  by using Eq. (\ref{eq:Bdiffeta}). We observe that, beyond the initial transient phase,  the mean-field solution is quite close to the one calculated from the DNS   in all of the cases and certainly acquires the correct decay rate. This serves as a validation of this approach. 

\par
\begin{figure}
\centering
\includegraphics[scale=0.35]{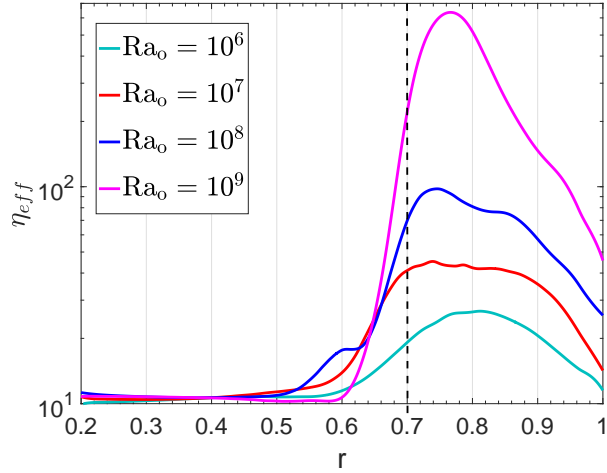}
\caption{\label{fig:Figure_11} Profiles of $\eta_{\rm eff}(r)$    for  Ra$_{\rm{o}}=10^6$,  Ra$_{\rm{o}}=10^7$,  Ra$_{\rm{o}}=10^8$ and Ra$_{\rm{o}}=10^9$.}
\end{figure}

With this mean-field framework, we can now shed light on the effect of turbulent convection on the large-scale dipole field in terms of two mean-field effects -- a pumping effect(${\gamma} \sim -(1/2)\nabla\eta_{\rm eff}$) and a turbulent diffusion effect ($\eta_{\rm eff}$) -- and try to establish how these depend on  Ra$_{\rm{o}}$.
 Figure \ref{fig:Figure_11} shows the enhanced turbulent diffusivity in the bulk of the CZ and in the overshoot region due to the stronger turbulent convective motions there. We see values significantly enhanced from the microscopic value of $1/{\rm Pm}=10$ everywhere except in the RZ, and generally increasing with Ra$_{\rm{o}}$. These profiles  strongly suggest that  the dipole magnetic energy in the DNS   decays faster than in the purely diffusive case  due to an enhanced turbulent diffusivity in the CZ that increases with   Ra$_{\rm{o}}$. 

\par
\begin{figure*}
\centering
\includegraphics[scale=0.5]{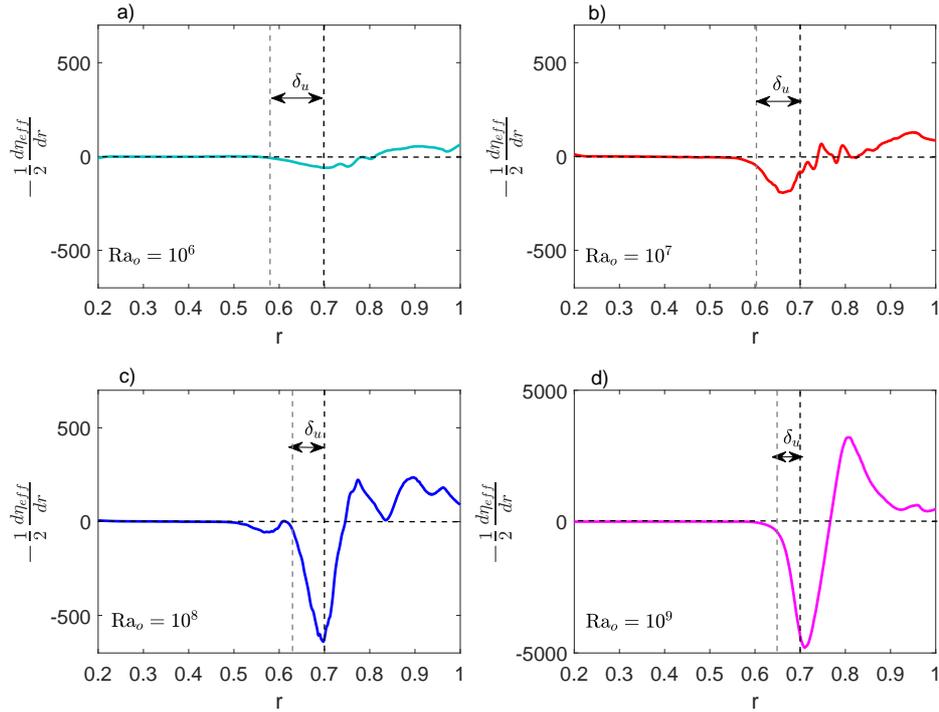}
\caption{\label{fig:Figure_12}  Profile of the diamagnetic pumping term $-\frac{1}{2}d\eta_{\rm eff}/dr$  against $r$ for  a) Ra$_{\rm{o}}=10^6$, b) Ra$_{\rm{o}}=10^7$, c) Ra$_{\rm{o}}=10^8$ and d) Ra$_{\rm{o}}=10^9$.}
\end{figure*}

 In Figure \ref{fig:Figure_12}, we plot the term ${\gamma}=-(1/2)d\eta_{\rm eff}(r)/dr$. This term can be interpreted as the diamagnetic  velocity that transports the  dipole magnetic field  down the gradient of turbulent intensity,   from the highly turbulent regions (i.e. the CZ and the overshoot region)   into the stable RZ below.
Consistent with the gradients seen  in Fig. \ref{fig:Figure_11}, we find  that this ${{\gamma}}$ pumping term  increases in magnitude with increasing Ra$_{\rm{o}}$, is maximal close to the    bottom of the CZ and drops to zero again outside of the overshoot layer (marked in the figure with the length-scale $\delta_{\rm u}$).    

\par
\begin{figure*}
\centering
\includegraphics[scale=0.7]{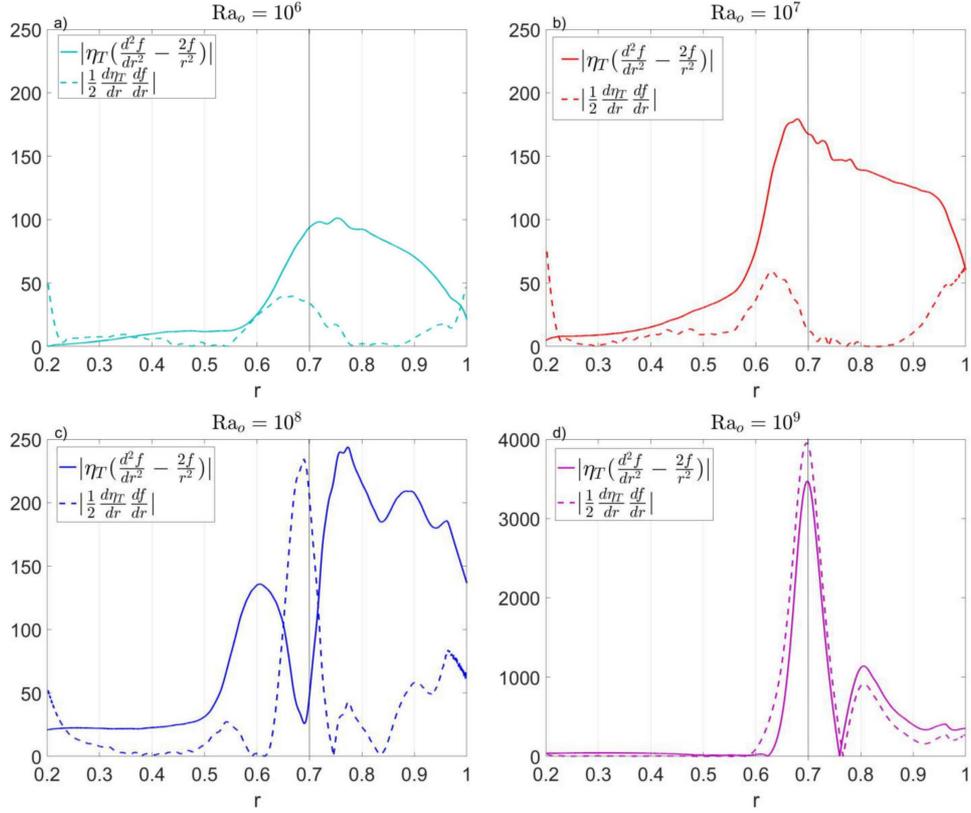}
\caption{\label{fig:Figure_13} Profile of the magnitude of the turbulent diffusion term (solid line) and the magnetic pumping term (dashed line) versus $r$ for a) Ra$_{\rm{o}}=10^6$, b) Ra$_{\rm{o}}=10^7$, c) Ra$_{\rm{o}}=10^8$ and d) Ra$_{\rm{o}}=10^9$.}
\end{figure*}

 To  understand the relative importance of  pumping (which transports the field inward) and  turbulent diffusion (which transports the field outward) in each case, in Figure \ref{fig:Figure_13}, we plot the magnitude of the full terms associated with the magnetic pumping $|(1/2)(d\eta_{\rm T}/dr)(df/dr)|$ and the turbulent diffusion $|\eta_{\rm T}({d^2f}/{dr^2}-{2f}/{r^2})|$, respectively, for the four values of the Rayleigh number. Both of these terms increase with increasing Ra$_{\rm{o}}$ as expected. For Ra$_{\rm{o}}=10^6$ and Ra$_{\rm{o}}=10^7$ turbulent diffusion is  larger than magnetic pumping. At Ra$_{\rm{o}}=10^8$, the pumping term  localized in the overshoot region becomes as strong as the maximum bulk turbulent diffusion. At Ra$_{\rm{o}}=10^9$, the behaviour in the overshoot zone dominates both terms, and the pumping term has finally exceeded the turbulent diffusion term. This can  explain  what we observe in Fig. \ref{fig:Figure_9}b, namely that magnetic pumping is  efficient enough to lead to the accumulation of the dipole field within and below the overshoot region.

\newpage

\section{Discussion}
\label{sec:discussion}
\subsection{Discussion of the results}
\par In this work, we  used DNS to elucidate   the dynamical interaction of overshooting convection with underlying magnetic fields.  To understand the  dynamics found in our simulations more intuitively,  we  compared our results to a mean-field model that contains both a turbulent diffusivity and a  pumping term related to the gradient of the turbulent diffusivity. From this comparison, we were able to extract  the turbulent diffusivity profile  $\eta_{\rm T}(r)$. Solving the forward mean-field problem (given in Eq. (\ref{eq:Bdiffeta}) subject to the same initial and boundary conditions as those used in the DNS) confirmed that this mean-field approach  indeed predicts reasonably well the overall decay rates of  the magnetic field observed in the DNS results. The fact that this model  works reasonably well  is perhaps somewhat surprising, since it  makes the simplistic  assumption that  the transport velocity (given by $\boldsymbol{\gamma}=-0.5\nabla\eta_{\rm T}$) is derived from  nearly isotropic  turbulence, which is certainly not the case for convection.
\par For this model to be useful as a predictive tool, it is necessary to understand the dependence of the turbulent diffusivity that underpins the model on parameters that might be known for stellar interiors.  Here we have kept the Prandtl number constant so our main concern is the dependence on the Rayleigh number. For that purpose, in Figure \ref{fig:Figure_14}, we plot the values of $\eta_{\rm CZ}$ (with their error bars -- see Appendix \ref{app:AppB} for more details) against Ra$_{\rm{o}}$, where we define $\eta_{\rm CZ}$ as
\begin{equation}
\label{eq:etacz}
\eta_{\rm CZ}=\dfrac{\int_{r_{\rm{t}}}^{r_{\rm{o}}}\eta_{\rm T}(r)r^2dr}{\int_{r_{\rm{t}}}^{r_{\rm{o}}}r^2dr}.
\end{equation}

\begin{figure}
\centering
\includegraphics[scale=0.32]{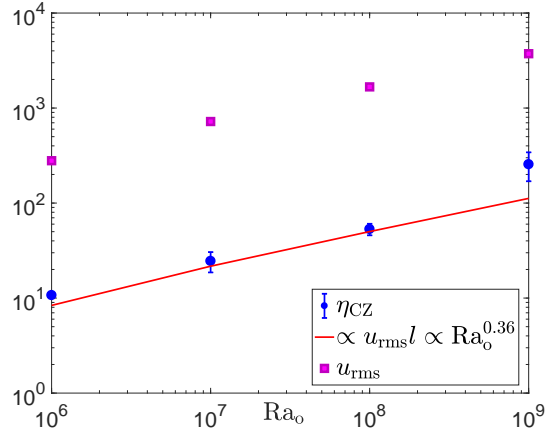}
\caption{\label{fig:Figure_14}  Plot of $\eta_{\rm CZ}$   versus the Rayleigh number along with the scaling  $u_{\rm rms}l$. The plot also shows the values of $u_{\rm rms}$ at each Ra$_{\rm o}$, where $u_{\rm rms}\propto{\rm Ra_o}^{0.36}$ \citep*[see][]{Korre}.}
\end{figure}
\noindent Since the Lorentz force is negligible, from a dimensional perspective we  expect that  $\eta_{\rm CZ}$ should scale as $\eta_{\rm CZ}\propto u_{\rm rms}l=$ Rm/Pm \citep[see, e.g.][]{CV91},  where $u_{\rm rms}$ is the non-dimensional rms velocity of the fluid extracted from the DNS, $l=0.3r_{\rm o}$ is the depth of the CZ, and Rm is the magnetic Reynolds number (see Table \ref{tab:MHDdata}) defined in terms of non-dimensional quantities as 
\begin{equation}
\label{eq:Rm}
{\rm Rm}=u_{\rm rms}l{\rm Pm}.
\end{equation}
We find that this scaling holds for the three lower Ra$_{\rm{o}}$ cases but not for the Ra$_{\rm o}=10^9$ simulation, even taking into account the errors in the measurement. The fact that the highest Ra$_{\rm o}$ case deviates from this scaling could be attributed  to two main possible reasons. First, it could be that the mean-field model is indeed a reasonable model of the fully nonlinear dynamics, but the correlations between the convection and the field that lead to $\eta_{\rm CZ}$ have actually changed. Second, it could be that the simplistic mean-field model adopted here fails at higher Ra$_{\rm o}$, omitting some dynamics that become important.  For example, the use of Eq. (\ref{eq:MF}) to model  $\boldsymbol{\gamma}$ is derived from  nearly isotropic  turbulence, and may become incorrect, or other anisotropic elements of the original $\boldsymbol{\alpha}$ tensor may become important.  Fitting of an incorrect mean-field model would in this case lead to inaccurate values of $\eta_{\rm CZ}$ that do not follow the expected scaling in Figure \ref{fig:Figure_14}.   In either case, the mismatch of the mean-field model and our  numerical results appears to indicate a substantial change in the dynamics at Ra$_{\rm{o}}=10^9$. It is important to further explore this new dynamical regime beyond Ra$_{\rm o}=10^9$, but unfortunately this  is presently not possible with PARODY due to computational constraints.

\subsection{Implications for  magnetic confinement and decay of the large-scale dipole field in the RZ}
\par Overall, our numerical results suggest that there is an increasing degree of ``confinement" of the dipole field to the stable RZ as the Rayleigh number increases, with a regime change  around Ra$_{\rm o}=10^9$, as described above. It is important to note that the confinement mechanism here is very different from that described theoretically by \citet{GM98} and exhibited in the models of \citet{AG13} and \citet{WB18}.  In these  models, the confining mechanism depends crucially on the presence of global rotation and is a slow, laminar process operating via large-scale meridional flows. Confinement is then viewed as a balance between downward advection and upward diffusion (see Section \ref{sec:intro}).  In our 3D simulations, which are without any rotation, there are no large-scale meridional flows and instead the confinement is achieved on a rapid advective time-scale by a balance between upward turbulent diffusion, and downward turbulent pumping. At the highest value of Ra$_{\rm o}$ achieved, the pumping is becoming sufficient to create a relatively confined state.  These results are very similar to what was found in \citet{Tobias2001} and in \citet{Tao98} who both studied turbulent transport of a mean-field in a Cartesian geometry via 3D and 2D numerical simulations, respectively. 
\citet{Tobias2001} performed a quantitative survey of the effect of $Q$ on the turbulent pumping of the magnetic field by varying $Q$ over five orders of magnitude.  They concluded that their results were very insensitive to this parameter.
\citet{Tao98} found that in the kinematic regime where the magnetic field is weak, the kinematic mean-field model works remarkably well in predicting the expulsion process of the field while they concluded that, in the dynamical regime, turbulent transport still takes place although not so efficiently as in the kinematic regime due to   suppression of the  effective turbulent diffusivity.  Therefore, following these prior studies indicating that there is overall little dependence of the dynamics on $Q$, we examined a low value of $Q$ to give this study the best chance of obtaining turbulent pumping, bearing in mind that we could explore the computationally harder dynamical regime in a future study given the necessary but currently not available  computational resources. 
Also, our explanation in terms of  mean-field theory is very similar to that which was proposed by \citet{KitsRud08}.   In that paper, only the forward problem was solved with an $\eta_{\rm T}(r)$ profile parametrised using an error function, transitioning between a high value in the CZ given by $\eta_{\rm CZ}$ and a much lower value in the RZ given by $\eta_{\rm in}$.  Despite the fact that the $\eta_{\rm T}(r)$ profile is quite different from the one we derive in our simulations, \citet{KitsRud08} also reported that poloidal field lines start to become confined at values of $\eta_{\rm CZ}/\eta_{\rm in} > 10$, which is commensurate with what we have for Ra$_{\rm o}=10^9$.  \citet{KitsRud08} report greater confinement at much higher ratios $\eta_{\rm CZ}/\eta_{\rm in}\sim 10^3-10^5$,  pointing to the tantalising need for simulations at higher Ra$_{\rm o}$. Based on their results, we might need $\eta_{\rm CZ}$ to be $10^2-10^4$ larger than what we currently have to obtain a more confined field in the RZ. Assuming that the traditional scaling of $\eta_{\rm CZ}\propto u_{\rm rms}l$, with $u_{\rm rms}\propto {\rm Ra}_{\rm o}^{0.36}$, still holds, this would require a Rayleigh number that is $\sim 10^6- 10^{12}$ times larger than the highest Rayleigh case we were able to simulate. 
\par It is possible that some factors may mitigate this issue however.  For example, it has long been suggested that topology can play a significant role in pumping \citep{DY74}. The simulations of \citet{Tobias2001} that showed relatively efficient pumping were carried out at only moderate Rayleigh numbers ($\sim 10^5$) but were performed in a  compressible fluid, where there is significant asymmetry in the convection.  Switching to more naturally asymmetric anelastic or compressible turbulence may  enhance turbulent pumping. 
\par It should be highlighted, as it was in \citet{Tobias2001} and found here, that pumping acts on large-scale (i.e. significantly larger than the advective scales) magnetic fields only. As such, there may be significant fluctuations in the field leading to a significant magnetic energy $\langle\B^2\rangle$, remaining in the overshoot layer, even though the mean field is confined below it. Similarly, small-scale magnetic fields (at the velocity scales or smaller) are constantly recirculating in the convection zone so the latter is by no means free of magnetic energy.
\par Finally, note that our data suggest that the decay rate ($\lambda$) of the dipole field  increases with increasing $\eta_{\rm CZ}$. 	 However,  in far more turbulent cases (where  the confinement would be more efficient),  we would expect that the decay rate of the large-scale dipole field would   depend less on what happened in the convective region  and  more on the  conditions within the CZ-RZ interface and the RZ, which would need to be modeled (although how remains to be determined).

\section{Conclusion and astrophysical implications}
\label{sec:conclusion}
\par Our results suggest that solar-type stars  with masses $\sim 1M_{\odot}$, which have thick convective envelopes with  extremely large Ra$_{\rm o}$, could possess primordial dipole magnetic fields that are fully confined in their stable region. By contrast, Ra$_{\rm o}$ rapidly decreases in higher-mass solar-type stars  with thinner outer convective regions, and are therefore less turbulent, and as a result, a mean poloidal field may not be able to remain confined in their RZ. 
\par We have discussed   two different possible mechanisms for confinement, namely the turbulent confinement process associated with convective overshooting motions operating on fast time-scales (studied in this paper) and the laminar confinement  associated with slow,  rotationally-driven meridional motions. It is very likely that both of these confinement  processes will play a role in the solar tachocline, where it has been shown that the presence of   a confined poloidal field in the stable region is needed to explain simultaneously the uniform rotation of the RZ and the thinness of the tachocline \citep[see, e.g.][]{GM98}. Also,   it is not unlikely that a model similar to the \citet{GM98} model may be able to account for the dynamical coupling of the core and envelope of RGB stars \citep[e.g.][]{Mosser2012}.
\par As discussed above, turbulent transport processes taking place in   these stars can only lead to the confinement of their large-scale field and small-scale fields will still be present in both the overshoot region and the convection zone, either as in these simulations, or by dynamo action.
Hence, it would be interesting to understand whether Ferraro's isorotation theorem can still persist in the same way under these more complicated conditions. Also, it is important to obtain better estimates of the amplitude of  the small-scale field residing in the tachocline. If this small-scale field is strong enough, it can significantly impact the conclusions of the \citet{GM98} model  which assumes a magnetic-free tachocline.  
\par Ultimately, our findings and conclusions suggest that it is of paramount importance to  now focus on the effects of the small-scale dynamo field existing in the vicinity of the CZ-RZ interface and understand its influence on the interior dynamics as it infiltrates the stable radiative zone from above.
These ideas will be further investigated  in the upcoming paper II  of this series of papers.

\section*{Acknowledgements} 
The authors thank Toby Wood for fruitful discussions. L.K. acknowledges support  from the George Ellery Hale Post-Doctoral Fellowship and from National Aeronautics and Space Administration (NASA) grant No. 80NSSC17K0008.
C.G. acknowledges support from the UK Natural Environment Research Council grant NE/M017893/1. This work was also partially  supported by National Aeronautics and Space Administration (NASA) Grant No. 80NSSC20K0602 (sub-award 62356550-145590).  The authors acknowledge the Texas Advanced Computing Center (TACC) at The University of Texas at Austin for providing HPC resources that have contributed to the research results reported within this paper (URL: http://www.tacc.utexas.edu). The initial hydrodynamic simulations were  run on the Hyades cluster at the University of California, Santa Cruz, purchased using the National Science Foundation (NSF) grant No. AST-1229745.

\section*{Data Availability}
The data underlying this article will be shared on reasonable request to the corresponding author. The PARODY-JA code is maintained by Julien Aubert and can be obtained upon request (\url{http://www.ipgp.fr/~aubert/Julien_Aubert,_Geodynamo,_IPG_Paris/Software.html}).
\bibliographystyle{mnras}

\bibliography{biblio}
 \appendix
 \section{SOLUTION OF THE DIFFUSION EQUATION} 
\label{app:AppA}

In this appendix, we derive the  solution of the  diffusion equation for the poloidal axisymmetric magnetic field $\B=\B_{\rm diff}$ given by
\begin{equation}
\label{eq:Bpol}
\dfrac{\partial \B_{\rm diff}}{\partial t}=-\nabla\times(\eta\nabla\times\B_{\rm diff}),
\end{equation}
where  $\eta$ could be a function of $r$. Following the work of  \citet{Chandra1956}, we can  express $\B_{\rm diff}$ in terms of the potential  $P(r,\theta,t)$  such that
\begin{equation}
\label{eq:Pot}
\B_{\rm diff}=\nabla\times(r\sin\theta P\hat{\bold{e}}_{\phi}).
\end{equation}
Substituting Eq. (\ref{eq:Pot}) into Eq. (\ref{eq:Bpol}) yields
\begin{equation}
\label{eq:Peq}
\dfrac{\partial^2 P}{\partial r^2}+\dfrac{4}{r}\dfrac{\partial P}{\partial r}+\dfrac{(1-\mu^2)}{r^2}\dfrac{\partial^2 P}{\partial\mu^2}-\dfrac{4\mu}{r^2}\dfrac{\partial P}{\partial\mu}=\dfrac{1}{\eta}\dfrac{\partial P}{\partial t},
\end{equation}
where $\mu=\cos\theta$.\\
We seek separable solutions of the form  $P(r,\mu,t)=R(r,t)G(\mu)$ 
so Eq. (\ref{eq:Peq}) yields separate equations for $R(r,t)$ and $G(\mu)$:
\begin{equation}
\label{eq:R}
\dfrac{\partial^2R}{\partial r^2}+\dfrac{4}{r}\dfrac{\partial R}{\partial r}-\dfrac{1}{\eta}\dfrac{\partial R}{\partial t}=\dfrac{\xi^2}{r^2} R,
\end{equation}
and
\begin{equation}
\label{eq:G}
(1-\mu^2)\dfrac{\partial^2 G}{\partial\mu^2}-4\mu\dfrac{\partial G}{\partial\mu}=-\xi^2 G.
\end{equation}
Equation (\ref{eq:G}) is the eigenvalue equation of the Gegenbauer polynomial $G_k^{3/2}(\mu)$,  with the eigenvalues given by $\xi_k^2=k(k+3)$, where $k\in \mathbb{N}$. \\
Hence, the full solution is given by \citep[see][]{Garaud99}
\begin{equation}
\label{eq:Psol}
P(r,\mu,t)=\sum_{k=0}^{\infty} A_k R_k(r,t)G_k^{3/2}(\mu).
\end{equation}
If we focus on the  dipole configuration, the only coefficient $A_k$ that is non-zero is for $k = 0$. The corresponding polynomial is simply $G_0^{3/2}(\mu) = 1$, with $\xi_0 = 0$. The remaining equation for $R_0(r,t)$  (Eq. (\ref{eq:R})) is:
\begin{equation}
\label{eq:R0}
\dfrac{\partial^2R_0}{\partial r^2}+\dfrac{4}{r}\dfrac{\partial R_0}{\partial r}-\dfrac{1}{\eta}\dfrac{\partial R_0}{\partial t}=0.
\end{equation}
The boundary conditions require that $R_0$ and $\partial R_0/\partial r$ be continuous at $r=1$, and that there is no singularity at the origin $r=0$, i.e. $R_0(0,t)=$ finite. Solutions of Eq. (\ref{eq:R0}) with these conditions are Bessel functions of fractional order. \\
The solution for $P(r,\theta,t)$ is then given by
\begin{equation}
\label{eq:Pfinal}
P(r,\theta,t)=\sum_{n=1}^{\infty}C_n r^{-3/2} J_{3/2}(n\pi r)\exp(-\eta n^2\pi^2 t),
\end{equation}
where $J_{3/2}$ is the Bessel function of order $3/2$, and where the coefficient $C_n$ is given by
\begin{equation}
C_n=\dfrac{\displaystyle\int_0^1P(r,\theta,0)r^{-3/2} J_{3/2}(n\pi r) r^4dr}{\displaystyle\int_0^1(r^{-3/2} J_{3/2}(n\pi r))^2r^4 dr}.
\end{equation}
Finally, the field components can be written in terms of $P$ (see Eq. (\ref{eq:Pot})) as
\begin{equation}
B_r=2\cos\theta P+\sin\theta\dfrac{\partial P}{\partial \theta},\quad B_{\theta}=-2\sin\theta P-r\sin\theta\dfrac{\partial P}{\partial r}.
\end{equation}

\vspace{5mm}
\section{CALCULATION OF ERRORBARS FOR $\eta_{\rm CZ}$} 
\label{app:AppB}
\par 
For the calculation of $\eta_{\rm T}$, we calculate the weighted-average $\bar{y}_{\rm dip}(r)$ (see Eq. (\ref{eq:ydip}))  from  $N$  available snapshots of the DNS at different times in the exponential decay phase. In order to obtain errorbars for $\eta_{\rm CZ}$, we split the  $N$ snapshots into $M$ subsets, each consisting of fewer than $N$  snapshots. Each one of these  subsets of snapshots corresponds to a time interval $\Delta t$ within the exponential decay phase of the simulation. We chose $M$ to ensure that $\Delta t$ contains  many convective turnover time-scales and more than one turbulent diffusion time-scale, in order for the calculated error to be physically reasonable. \\
Following the same process as described in Section \ref{sec:mft} for the calculation of $\eta_{\rm T}(r)$,  we  obtain a profile of $\eta_{{\rm{T}}_i}$ in each individual subset ($i=1,...,M$).
From that, we  then calculate $\eta_{\rm rms}(r)$ as
\begin{equation}
\label{eq:etarms}
\eta_{\rm rms}(r)=\left(\dfrac{1}{M}\sum_{i=1}^{M}(\eta_{{\rm{T}}_i}(r)-\eta_{\rm T}(r))^2\right)^{1/2}.
\end{equation}
Finally,  the errorbar for $\eta_{\rm CZ}$ can be found by using Eq. (\ref{eq:etacz}) where we have replaced $\eta_{\rm T}$ with  $\eta_{\rm T}(r)\pm\eta_{\rm rms}(r)$.

\bsp	
\label{lastpage}
\end{document}